\documentstyle[12pt,epsf]{article}
\def\period{\, .}
\def\comma{\, ,}
\def\del        {  \partial  }
\def\half       {  {1\over 2}  }

\def\defint#1#2 {  \int_{#1}^{#2}  }
\def\rootof#1   {  \left( #1 \right)^{1/2}  }
\def\deldel#1   {  {\partial\over \partial #1}  }

\def\abs#1      {  \vert #1 \vert  }
\def\ie         {  {\it i.e.}      }
\def\evalat#1   {  \left\vert_{#1} \right. }
\def\e          { {\rm e}  }

\def\lsim    {\lower .65ex \hbox{\ $\stackrel{<}{\sim}$\ } }
\def\gsim    {\lower .65ex \hbox{\ $\stackrel{>}{\sim}$\ } }

\def\calL       { {\cal L} }

\def\calO       { {\cal O} }

\def\vecii#1#2      {  \left(\begin{array}{c}#1\\#2\end{array}\right)  }
\def\veciii#1#2#3   {  \left(\begin{array}{c}#1\\#2\\#3\end{array}\right)  }
\def\veciv#1#2#3#4  {  \left(\begin{array}{c}#1\\#2\\#3\\#4
                                 \end{array}\right)  }
\def\vecv#1#2#3#4#5 {  \left(\begin{array}{c}#1\\#2\\#3\\#4\\#5
                                 \end{array}\right)  }

\def\matrixii#1#2#3#4            {  \left(\begin{array}{cc}#1&#2\\#3&#4
                                       \end{array}\right) }
\def\matrixiii#1#2#3#4#5#6#7#8#9 {  \left(\begin{array}{ccc}#1&#2&#3\\
                                     #4&#5&#6\\#7&#8&#9\end{array}\right)  }
\def\mativ#1#2#3#4               {  \left(\begin{array}{cccc}
                                       #1\\#2\\#3\\#4\end{array}\right) }
\def\matv#1#2#3#4#5              {  \left(\begin{array}{ccccc}
                                     #1\\#2\\#3\\#4\\#5\end{array}\right)  }
\def\eqabegin         {  \begin{eqnarray}  }
\def\eqaend           {  \end{eqnarray}  }
\def\nn               {  \nonumber  }
\def\bracetwo#1#2     {  \left\{ \begin{array}{l} #1 \\ #2 \end{array}
                         \right.  }
\def\bracetwocases#1#2#3#4  {   \left\{ \begin{array}{ll} #1 &
                                 \qquad #2 \\
                                 #3 & \qquad #4 \end{array} \right.  }
\def\bracebegin#1     {  \left\{ \begin{array}{#1}   }
\def\braceend         {  \end{array}\right.   }
\def\parn              {  \par\noindent }
\def\parbigskip        {  \par\bigskip  }
\def\parmedskip        {  \par\medskip  }
\def\parsmallskip      {  \par\smallskip  }
\def\parbigskipn        {  \par\bigskip\noindent  }
\def\parmedskipn        {  \par\medskip\noindent  }

\def\parag#1           {\paragraph{#1} \mbox{ }\parmedskip\noindent}
\def\boxit#1#2      {  \vbox{\hrule\hbox{ \hskip -4.1pt \vrule\kern3pt
                       \vbox
                    {  \hsize #1 \strut\kern3pt #2 \kern3pt\strut  }
                       \kern3pt  \vrule} \hrule  } }
\def\centerbox#1#2  {  \mbox{  }\par\bigskip  \hfil \boxit{#1}{#2} \hfil
                       \par\bigskip\noindent }
\def\rightbox#1#2   {  \hfill\boxit{#1}{#2}  }
\def\leftbox#1#2    {  \boxit{#1}{#2}  }
\def\fullbox#1      {  \boxit{\textwidth}{#1}  }
\def\trianglemap#1#2#3#4#5#6  {   {\large $$ \begin{array}{rcl} #1\!\!\!
                                  &{\stackrel{{\scriptstyle #2}}{
                              \longrightarrow   }}&\!\!\!  #3 \\
                            { } & {\scriptstyle #4}\!\!\!\searrow \quad
                                \swarrow \!\!\!{\scriptstyle #5}& { } \\
                                  { } & #6 & { } \end{array} $$ }    }
\def\squaremap#1#2#3#4#5#6#7#8    { {\large $$ \begin{array}{ccc}#1 &
                   \stackrel{{\scriptstyle #2}}{\longrightarrow} & #3 \\
                     {\scriptstyle #4}\!\downarrow & { } & \downarrow \!
                     {\scriptstyle #5}\\ #6 &\!\!
                      \longrightarrow_{{ }_{\!\!\!\!\!\!\!\!\!\!\!
                      {\scriptstyle #7}}}    &#8 \end{array} $$ }   }
\def\righttrianglemap#1#2#3#4#5#6  {  {\large $$ \begin{array}{rcl}
                 #1\!\! & \stackrel{{\scriptstyle #2}}{\longrightarrow}
                      & #3 \\  { }&\!\!{\scriptstyle #4}\!\!\searrow
                      & \downarrow \!\!{\scriptstyle #5}\\
                      { }&{ }& #6 \end{array} $$ }   }
\def\rightfigspacebegin  {  \par\noindent\begin{minipage}[t]{10cm}  }
\def\rightfigspaceend    {  \end{minipage}\par\noindent  }
\def\leftfigspacebegin   {  \par\noindent
                             \hspace*{10cm}\begin{minipage}[t]{6cm} }
\def\leftfigspaceend     {  \end{minipage}\par\noindent  }
\def\titleandfile#1#2   {  \begin{center}{\Large\bf #1}\end{center}
                            \par\begin{flushright} #2 \end{flushright}  }
\def\msection#1      {  \begin{center} \section{#1} \end{center}   }
\def\nsection#1      {  \let\boldface\bf \def\bf{} \section{#1}
                           \let\bf\boldface   }
\def\mnsection#1     {  \begin{center} \nsection{#1} \end{center}  }
\def\capsection#1    {  \let\boldface\bf \def\bf{\sc} \section{#1}
                           \let\bf\boldface   }
\def\mcapsection#1   {  \begin{center} \capsection{#1} \end{center} }
\def\sectionnumbering { \setcounter{equation}{0}
         \renewcommand{\theequation}{\arabic{section}.\arabic{equation}}}
\newcommand{\nullify}[1]{}
\def\ff{{1\over \gamma^2}}
\def\delmu{\del_\mu}
\def\delnu{\del_\nu}
\def\delplus{\del_+}
\def\delminus{\del_-}
\def\xplus{{x^+}}
\def\xminus{{x^-}}
\def\X{X}
\def\ep{\epsilon}

\def\d{\Delta}
\def\ghat{\hat{g}}

\def\qphi{q_{,\phi}}
\def\Vphi{V_{,\phi}}
\def\vphi{v_{,\phi}}

\def\papertitlepage{\baselineskip 3.5ex \thispagestyle{empty}}
\def\Title#1{\baselineskip 1cm \vspace{1.5cm}\begin{center}
 {\Large\bf #1} \end{center}
\vspace{0.5cm}}
\def\Authors#1{\begin{center} {\it #1} \end{center}}
\def\Abstract{\vspace{1.0cm}\begin{center} {\large\bf Abstract}
           \end{center} \parbigskip}
\def\Komabanumber#1#2#3{\hfill \begin{minipage}{4.2cm} UT-Komaba #1
              \parn #2
              \parn #3 \end{minipage}}
\renewcommand{\thefootnote}{\fnsymbol{footnote}}
\renewenvironment{thebibliography}{\pagebreak[3]\par\vspace{0.6em}
\begin{flushleft}{\large \bf References}\end{flushleft}
\vspace{-1.0em}

\begin{enumerate}\if@twocolumn\baselineskip=0.6em\itemsep -0.2em
\else\itemsep -0.2em\fi\labelsep 0.1em}{\end{enumerate}}
\begin{document}
\papertitlepage
\vspace*{0cm}
\Komabanumber{94-16}{hepth@xxx/9409179}{September 1994}
\Title{A Unified Approach to  Solvable Models of
 Dilaton Gravity in Two-Dimensions
 Based on Symmetry }
\Authors{{\sc
Y.~Kazama,
\footnote[2]{e-mail address:\quad
kazama@hep3.c.u-tokyo.ac.jp}
\ \ Y.~Satoh
\footnote[3]{e-mail address:\quad
ysatoh@hep1.c.u-tokyo.ac.jp}  \\
and \\
A. Tsuchiya
\footnote[4]{e-mail address:\quad
tsuchiya@hep1.c.u-tokyo.ac.jp}  \\
}
\vskip 3ex
 Institute of Physics, University of Tokyo, \\
 Komaba, Meguro-ku, Tokyo 153 Japan \\
  }
\vspace{-0.5cm}
\Abstract
A large class of solvable models of dilaton gravity in two space-time
 dimensions, capable of describing black hole geometry,  are analyzed
in a unified way as non-linear sigma models possessing  a special
symmetry. This symmetry, which can be neatly
 formulated in the target-space-covariant manner, allows one to
 decompose the non-linearly interacting dilaton-gravity system into a
free field and a field satisfying the Liouville equation with
in general non-vanishing  cosmological term. In this formulation, all
the existent models are shown to fall into the category with vanishing
cosmological constant. General analysis of the space-time structure
 induced by a matter shock wave is performed and new
 models, with and without the cosmological term, are discussed.
\baselineskip=0.7cm
\parbigskip
PACS number(s): 04.06.+n
\newpage
\baselineskip=0.7cm
\section{Introduction}
\renewcommand{\thefootnote}{\arabic{footnote}}
\sectionnumbering
Everyone would agree that quantum physics of black holes is one of
 the most fascinating and challenging subjects in
theoretical physics. It is the arena in which the two fundamental
 principles of modern physics, namely   quantum mechanics and
  general relativity, are deeply interwoven to produce effects which
 defy our conventional wisdom. The ultimate fate of a black hole
 emitting Hawking radiation, the possible loss of quantum coherence
 across the horizon, and the statistical meaning of the black hole
 entropy are among such conundrums. What makes these questions
 difficult to answer is, among other things, that they
 require us to treat the matter and the gravitational
 degrees of freedom on the same footing, including their
quantization.  Unfortunately, the present state of the art is not advanced
enough  to do this in a realistic four dimensional setting and we need to
 resort to some simplified models in low dimensions. \parsmallskip
Even in low dimensions, the task is far from trivial.  Although interesting
 in many respects, non-linear sigma model description of string theory
 with a black hole in the target space \cite{BRS,Wi} is
not capable of
describing  the all important back reaction. Recent attempts using
 suitable  matrix models \cite{JY}  should  in principle be free
of such limitations  but there the space-time interpretation
 presents difficulty. \parsmallskip
  Along a more conventional field theoretical line,  an attractive
 string-inspired dilaton gravity model was presented by Callan, Giddings,
 Harvey and  Strominger (CGHS) \cite{CGHS} and has aroused a lot of
 interest as it contains black hole solutions and is classically exactly
solvable \cite{RS1}-\cite{PS}. With this model, the authors showed
 that one can partly
incorporate the effects of back reaction in the limit of large number $N$ of
 matter fields in the form of anomaly-induced effective action and
 demonstrated the existence of  Hawking radiation.  The problem was,
 however, that this large $N$ approximation broke down for small
 black hole mass and hence the fate of the black hole
 could not be traced to the end. Furthermore, since the dilaton-gravity
 sector was not really quantized,  the question of the possible loss of
 quantum coherence could not be addressed.  It was clear that
 some solvable extensions or modifications were called for.  \parsmallskip
Since then, a number of  such solvable models of CGHS type have been
 devised and analyzed from various points of view
\cite{Al}, \cite{BC}, \cite{RST}, \cite{HKS}-\cite{KS2}, \cite{VV},
\cite{HA}-\cite{Mi1}.
In addition to being
 exactly solvable at the semiclassical level,  these models are
 in principle amenable to full quantum treatment  albeit  with certain
 caveat about the range of the fields and the choice of the functional
 measure.  Indeed for the simplest of these models, exact quantization
 has been performed and some physical consequences have been
 extracted \cite{HKS}-\cite{KS2}, \cite{VV}.
Although satisfactory answers to the
 aforementioned questions concerning  the fate of the black hole etc.
 are yet to be sought for, these developments have demonstrated the
 usefulness of such solvable models of dilaton gravity.
( For pedagogical reviews of the CGHS model and the subsequent
 developments, see \cite{HS}, \cite{Gi}, \cite{Ka}.)
\parsmallskip
The main objective of the present article is to characterize and clarify
the structure of a large class of such solvable models from a unified
 point of view.  As has been emphasized in the literature
 \cite{GS,Al}, there is an immense
  freedom in the possible form of the action for  dilaton gravity models in
 two dimensions due to the dimensionless nature of the dilaton
 field. Namely, when one takes  into account the effects of  functional
 measure and renormalization,  the general form of the action can take
 that of a non-linear sigma model formally similar to the string
theory in curved space.  Of course in order to be regarded as models
 of dilaton gravity, they must satisfy such requirements as
two-dimensional general  covariance  and conformal invariance,
 but this still allows for infinite possibilities for the choice of the
 form of the \lq\lq target  space metric", the coupling to the background
 scalar curvature etc..  This freedom is also reflected in the variety of
 solvable models  proposed.  What we wish to do is to
 capture the characteristic feature common to all such solvable models and
 thereby treat  them in a systematic way. \parsmallskip
A conspicuous feature of  the solvable models so far proposed
 is that  by some  field transformations one can reduce
 the models down to those of massless free fields. These transformations
are however generally quite complicated and require
 certain amount of ingenuity to be discovered.
 What we shall show in this paper is that behind the existence of
 such transformations and hence the solvability of the models is
 in fact a powerful symmetry, which is an extensive generalization of
 the one noted in the specific model of Russo, Susskind and Thorlacius
 \cite{RST}.  This symmetry, which can be neatly formulated in the target-space
 covariant manner in the general non-linear sigma model formulation
 mentioned above,  guarantees the existence of at least
one free field, which we call   $F(x)$, in the dilaton-gravity sector, and
 further dictates  that  the field  orthogonal to $F(x)$, which will be
 called $V(x)$, satisfies a Liouville equation with in general non-vanishing
 cosmological constant.  It turned out that all the solvable models
 so far proposed fall into this scheme with vanishing  cosmological
 constant.  This means that  not only do we have a unified framework to discuss
 existent  models but also are able to treat a new class of solvable models
with  a genuine Liouville type field.  Indeed, after deriving some
 formulas to study
 the space-time structure  induced by a matter shock wave in a general
 manner, we shall analyze new   models,
with and without the cosmological term. \parsmallskip
Let us  briefly indicate the organization of the rest of this article:
In Sec.2, after specifying the non-linear sigma model we deal with,
 we describe the symmetry which allows us to decompose the system
 into a free field and a Liouville field. Condition of conformal invariance
 further determines the structure of the background charge terms and
 the system is completely solved in terms of two free fields.  In Sec.3,
 we apply this  formalism to a general class of dilaton gravity models.
 The condition of solvability following from the aforementioned symmetry
 takes the form of a differential equation relating various functions
specifying the model and it is solved to give an explicit  relation
among them.  From this formula we immediately see that  existing
 models are realized as special cases of our general treatment.
 We then go on in Sec.4 to investigate the space-time structure of
 the models  induced by a matter shock wave. By studying  the
 singularity structure and the asymptotic form  of the scalar curvature,
  we are lead to analyze a number of new models which exhibit
 a variety of  physical  behavior.
Finally, Sec.5 is devoted to discussions concerning
 full quantization and on possible implication of
our model as a model of string theory.
\section{ Sigma Model with a Symmetry}
\sectionnumbering
\subsection{ The Action }
As was recalled in the introduction,  due to the ambiguities in the
choice of the functional measure and to the uncertainties in the
renormalization counter terms,  a systematic study of  dilaton gravity
 models must deal with an action of  a non-linear sigma model type.
 In this paper, for definiteness  we shall focus on  such an  action of the
 form
\eqabegin
 S &=& \ff \int d^2x \sqrt{-\ghat}\, \left[ \ghat^{\mu\nu}
 \delmu \X^i G_{ij}(\X) \delnu\X^j + Q(\X)\hat{R}
 +\Lambda \e^{V(\X)} \right] \comma \label{eqn:action}
\eqaend
where $\gamma$ and $\Lambda$ are constants and the \lq\lq target
space coordinates"
$\X^i(x), (i=1,2) $ are to be identified later with the dilaton field
$\phi(x)$ and the conformal factor $\rho(x)$ in a certain conformal
gauge. We shall take the signature of the world-sheet metric to be
$( - + )$.
 \parsmallskip
Let us first study the properties of this action for the flat
 (world-sheet) reference metric $\ghat_{\mu\nu} = \eta_{\mu\nu}$,
 namely
\eqabegin
S_{flat} &=& \ff \int d^2x  \left[ \delmu \X^i
G_{ij} \del^\mu \X^j + \Lambda \e^V \right] \label{eqn:faction}\period
\eqaend
Under a general variation $\delta X^i$, the Lagrangian changes by
\eqabegin
\gamma^2 \delta \calL
&=&  2\del^\mu \left( \delmu \X^i G_{ij}  \delta\X^j\right)
-2\left( \Box \X^k + \Gamma_{ij}^k\delmu\X^i\del^\mu \X^j\right)G_{kl}
\delta\X^l
 +\Lambda V_{,k}\delta\X^k \e^V \comma \label{eqn:lagvar}
\eqaend
where $\Gamma^k_{ij}$ is the usual Christoffel symbol constructed from
 the target space metric $G_{ij}$, $\Box$ is the world-sheet Laplacian,
 and we have used the abbreviation
 for the target space derivative $V_{,k} \equiv \del V /\del X^k$.
{}From this expression, we can immediately read off the equations of motion
\eqabegin
 \Box \X^k &=& -\Gamma^k_{ij}\delmu\X^i\del^\mu\X^j + \half
\Lambda V^{,k} \e^V  \label{eqn:eqmotion} \period
\eqaend
%
\subsection{ Symmetry of the Model }
What will be of utmost importance is the observation that the
 system possesses a symmetry if the variation $\delta X^k$
 satisfies the following two conditions:
\eqabegin
&& (i)\qquad V_{,k} \delta\X^k = 0 \comma \label{eqn:symi}\\
&& (ii) \qquad \nabla_i\delta\X_j = \del_i \delta\X_j -\Gamma_{ij}^k
\delta\X_k =0 \comma \label{eqn:symii}
\eqaend
where $\delta\X_j \equiv G_{jl}\delta\X^l$. Indeed, for such variations,
 the last term in (\ref{eqn:lagvar}) vanishes, while the second term
 can be rewritten as
\eqabegin
&& -2\left( \Box \X^k \delta\X_k + \Gamma_{ij}^k\delmu\X^i\del^\mu
\X^j\delta\X_k \right) \nn\\
&& = -2\left( \Box \X^k \delta\X_k+\del_i\delta\X_j
\delmu\X^i\del^\mu \X^j\right)  \nn\\
&& = -2\del^\mu \left( \delmu\X^k \delta\X_k \right) \comma
\eqaend
which precisely cancels the first term. Thus we find
 $\delta \calL =0$ and we have a symmetry. \parsmallskip
In two dimensions, the condition $(i)$ can be solved uniquely (up to
 a constant):
\eqabegin
\delta\X^k &=& {\epsilon^{kl}\over \sqrt{-|G|}} V_{,l} \comma
\eqaend
where $|G|$ is the determinant of $G_{ij}$ and $\epsilon^{kl}$
 is the usual antisymmetric matrix with $\ep^{12}= - \ep_{12} = 1$.
Then substitution into $(ii)$ yields a condition for $V(X)$:
\eqabegin
\nabla_i \nabla_j V &=& 0 \period \label{eqn:ddv}
\eqaend
This means that $V_{,k}$ is a target space Killing vector, and moreover,
from  $0 = \left[\nabla_i, \nabla_j\right] V_{,k} =
 R_{lkij}V^{,l}$, we deduce that the two-dimensional target space must
be flat. \parsmallskip
An important consequence of this symmetry is that it guarantees
 the existence of a free field. Let $F(X)$ be a function of $X^i$ and
 apply to it the world-sheet Laplacian. Using the equations of motion
 for $X^i$ we get
\eqabegin
 \Box F(\X) &=& \Box\X^k F_{,k} + \delmu\X^i\del^\mu\X^j F_{,ij} \nn\\
 &=& \nabla_i\nabla_j F  \delmu\X^i\del^\mu\X^j
+\half \Lambda F_{,k}V^{,k}\e^V \period \\
\eqaend
This vanishes if
\eqabegin
 \nabla_i\nabla_j F &=& 0 \comma \\
 F_{,k}V^{,k} &=& 0 \period
\eqaend
The second condition is solved by
\eqabegin
 F_{,k} &=& - \sqrt{-|G|}\, \epsilon_{kl} V^{,l} \label{eqn:Fdef} \comma
\eqaend
which automatically satisfies the first condition as well since
$\nabla_i\nabla_j V=0$. To assure that the function $F(X)$ itself
 exists, we must check the integrability condition, \ie if
$ \epsilon^{kl} \nabla_k F_{,l} =0$ holds.
After a simple calculation we get
\eqabegin
 \epsilon^{kl} \nabla_k F_{,l} &=& - \sqrt{-|G|} \nabla_m V^{,m}
 =0 \comma
\eqaend
again due to $\nabla_i\nabla_j V=0$. This establishes
  that $F(X)$ defined by Eq.(\ref{eqn:Fdef}) is a free field.
\parsmallskip
In fact it is easy to show that $F(X)$ is directly related to
 the Noether current associated with the symmetry.
Recall that $\delta \calL=0$ identically under
 the symmetry variation. Thus the Noether current takes a
 simple form
\eqabegin
j_\mu &=& \delmu X^i G_{ij} \delta X^j \nn \\
 &=& \delmu X^i G_{ij} {\epsilon^{jk} \over \sqrt{-|G|} } V_{,k}
\label{eqn:j} \period
\eqaend
By making use of $G_{ij}\epsilon^{jk}/\sqrt{-|G|}
 = \sqrt{-|G|} \epsilon_{ij} G^{jk}$, we find
\eqabegin
 j_\mu &=& \delmu X^i \sqrt{-|G|}\, \epsilon_{ij} V^{,j} \nn\\
 &=&-\delmu F \period
\eqaend
Therefore the current conservation and the free field equation
 for $F(X)$ are identical.  This type of symmetry was previously
noticed in a particular model of dilaton gravity \cite{RST}
\footnote{In this model, the variation and the current take the form
$ \delta \phi = \delta \rho = \ e^{2\phi}$ and
$ j^\mu = \del^\mu( \phi - \rho ) $
respectively in their notation.}.
We now
 have its generalization and will demonstrate its power as we develop
 our general treatment. \parsmallskip
Let us now try to express the action (\ref{eqn:faction}) in terms
 of $F(x)$ and $V(x)$. To do this, compute $j_\mu j^\mu$ with $j_\mu$
 given in (\ref{eqn:j}). Using the identity
$\epsilon^{ij}\epsilon_{mn} =
- (\delta^i_m \delta^j_n  -\delta^i_n \delta^j_m) $, we get
\eqabegin
 j_\mu j^\mu &=& -\left( \delmu X^i G_{ij} \del^\mu X^j \right)
 V^{,k} V_{,k} + \delmu V \del^\mu V \period
\eqaend
When $\nabla_i \nabla_j V =0$ holds, it is easy to see that  the
 squared norm of the vector $V_{,k}$
\eqabegin
 \d &\equiv & V^{,k} V_{,k}
\eqaend
 is a constant.  Thus,  with $j_\mu =- \delmu F$,
 we have the action in the form
\eqabegin
S_{flat} &=& \ff \int d^2x  \left[ \delmu \X^i
G_{ij} \del^\mu \X^j + \Lambda \e^V \right] \nn\\
 &=& \ff \int d^2x  {1\over \d}
 \left[ -\delmu F \del^\mu F + \delmu V \del^\mu V
+ \Lambda \d \e^V \right] \label{Sflat} \period
\eqaend
We see that, as announced in the introduction,
 the system decomposes into a free field $F(x)$ and
 a field $V(x)$ orthogonal to it which satisfies  the Liouville equation
\eqabegin
 \Box V &=& \half \Lambda \d \e^V \period
\eqaend
Its general solution is well-known \cite{Li} and can be expressed
 in terms of a free field $\psi$  in the form
\eqabegin
 V &=& \psi -2\ln Y \comma \label{eqn:solV} \\
 Y &=& 1+{\Lambda \d \over 16} A(\xplus) B(\xminus) \comma
 \label{eqn:solY}
\eqaend
where $A(\xplus) $ and $B(\xminus)$ are defined by
\eqabegin
 \delplus A (\xplus) \delminus B(\xminus) &=& \e^\psi
\period \label{eqn:AB}
\eqaend
%
\subsection{ Conditions for Conformal Invariance }
In order for a sigma model to describe a world-sheet gravity theory,
 it is necessary that it has the conformal symmetry. This will impose
 conditions for the field $Q(X)$ coupled to the background curvature
 $\hat{R}$. \parsmallskip
By varying (\ref{eqn:action}) with respect to the reference metric
$\ghat^{\mu\nu}$,
 one obtains the energy-momentum tensor $T^{DG}_{\mu\nu}$
( in a suitable normalization ) for the dilaton-gravity sector.
For flat metric $\ghat_{\mu\nu} = \eta_{\mu\nu}$,
$T^{DG}_{\mu\nu}$ and its trace take the form
\eqabegin
 T^{DG}_{\mu\nu} &=& -\delmu\X^i G_{ij} \delnu\X^j +\half \eta_{\mu\nu}
 \left( \del_\alpha\X^i G_{ij} \del^\alpha\X^j\right) \nn\\
 && - \left( \eta_{\mu\nu}\Box -\delmu\delnu\right) Q
 +\half\eta_{\mu\nu} \Lambda \e^V \comma\\
{ T^{DG}}_\mu ^\mu &=& -\Box Q +\Lambda\e^V
 = -\Box\X^i Q_{,i} - \delmu\X^i\del^\mu
 \X^j Q_{,ij} +\Lambda\e^V \period
\eqaend
With the use of   the equations of motion (\ref{eqn:eqmotion}), the
 condition for the vanishing of ${T^{DG}}_\mu^\mu$ can be expressed as
\eqabegin
 \half \Lambda \left( Q^{,k}V_{,k}-2\right)\e^V
 &=& - \nabla_i \nabla_j Q
  \delmu \X^i \del^\mu\X^j \period
\eqaend
Since we can change $\delmu\X^i$ while keeping
 the values of $X^i$ fixed, we must have
\eqabegin
 Q^{,k}V_{,k} &=& 2 \comma \\
 \nabla_i \nabla_j Q &=& 0 \period
\eqaend
The second condition tells us that $Q_{,k}$ is a  Killing vector
with a constant norm
and hence it must be expressible as a linear combination of two other
 ( orthogonal ) Killing vectors  $F_{,k}$ and $V_{,k}$.
Writing this relation in the form
\eqabegin
 F_{,k} &=& c_V V_{,k}+ c_Q Q_{,k} \comma   \nn
\eqaend
and contracting with $V^{,k}$, we immediately find
\eqabegin
 F_{,k} &=& c_V \left( V_{,k} -\half \d  Q_{,k} \right)
 \period\label{eqn:FVQ}
\eqaend
The constant $c_V$ can be expressed in terms of $\d$ and the squared
 norm $Q_{,k}Q^{,k}$: First from (\ref{eqn:Fdef}) we find $F_{,k}F^{,k}
 = -\d$.  On the other hand, the same quantity can be computed from
Eq. (\ref{eqn:FVQ}) above. Comparing them we get
\eqabegin
 c_V &=& {1 \over \sqrt{1-{1\over 4} \d  Q_{,k}Q^{,k} } } \period
\label{eqn:cVQ}
\eqaend
The value of $Q_{,k}Q^{,k}$ will be specified for dilaton gravity models
 in the next section. \parsmallskip
  Solving the equation (\ref{eqn:FVQ})
 for $Q$, we finally obtain the expression for
 the full action (\ref{eqn:action}) in terms of $F$ and $V$:
\eqabegin
 S &=& \ff \int d^2x \sqrt{-\hat{g}}\, {1\over \d} \Biggl[ -\hat{g}^{\mu\nu}
 \delmu F \delnu  F + \hat{g}^{\mu\nu} \delmu V \delnu V
 + \Lambda \d  \e^V \nn\\
 && \qquad + \left(2V -{2\over c_V} F\right) \hat{R} \Biggr]
\label{eqn:FVaction} \period
\eqaend
The energy-momentum tensor following from this action is
\eqabegin
 T^{DG}_{\pm\pm} &=& T^F_{\pm\pm} + T^V_{\pm\pm} \comma\nn\\
 T^F_{\pm\pm} &=& {1\over \d} \left( (\del_\pm F)^2 - {2\over c_V}
 \del_\pm^2 F \right) \comma \label{eqn:EM} \\
T^V_{\pm\pm}&=& {-1\over \d} \left( (\del_\pm V)^2 - 2 \del_\pm^2 V \right)
= {-1\over \d} \left( (\del_\pm \psi)^2 - 2 \del_\pm^2 \psi\right)
\period \nn
\eqaend
These expressions appear to be singular when the $\d \rightarrow 0$,
 \ie when $V$ becomes a free field. In fact from (\ref{eqn:FVQ}) one
 sees that $F$ and $V$ become degenerate in this limit.
 This however is not a problem. From (\ref{eqn:cVQ}) we see that
  $c_V = 1 + \calO(\d)$ in that limit
 and hence $F-\psi$ will be of $\calO(\d)$, as is easily seen from
 (\ref{eqn:FVQ}) and the structure of $V$ in (\ref{eqn:solY}).  With this
 in mind, we define, for general $\d$,  a new free field $\chi$
 by the equation
\eqabegin
 F &=& {1\over c_V}\psi +{c_V\d \over 2} \chi \label{eqn:defchi} \period
\eqaend
  Then after a simple calculation we get
\eqabegin
 T^{DG}_{\pm\pm} &=& \del_\pm \psi \del_\pm \chi
+{c_V^2\d \over 4} (\del_\pm\chi)^2 -  \del^2_\pm \chi
 - {1\over 4}Q_{,k}Q^{,k}
\left( (\del_\pm \psi)^2 -2\del^2_\pm \psi \right)
\period \label{eqn:EMpc}
\eqaend
This expression is
 completely regular as $\d \rightarrow 0$ and the system will be
described by a pair of free fields $\psi$ and $\chi$.
\section{ General Analysis of Solvable Models of Dilaton Gravity}
\sectionnumbering
\subsection{ The Model }
Having formulated a sigma model with a special symmetry, we now
 apply it to dilaton gravity models in a unified
 manner.  A large class of such models can be represented by the
action of the form
\eqabegin
 S &=& \ff\int d^2x \sqrt{-g}\, \left[ g^{\mu\nu}K(\phi)\delmu
\phi \delnu \phi + q(\phi) R + \Lambda \e^{v(\phi)}\right]
\comma \label{eqn:dgaction}
\eqaend
where $\phi$ is the dilaton field.
In order to be able to deal with possibilities of various conformal
 frames, we shall first make a conformal transformation of the metric
 by a factor $\e^{2\omega(\phi)}$ and then take a conformal gauge.
 This amounts to the combined  transformation
\eqabegin
 g_{\mu\nu} &=& \e^{2(\rho+\omega(\phi))} \ghat_{\mu\nu} \period
\eqaend
Then  the curvature scalar takes the form
\eqabegin
  R &=& \e^{-2(\rho+\omega)}\left( \hat{R} -2\ghat^{\mu\nu}
 \hat{D}_\mu \hat{D}_\nu (\rho+\omega) \right) \period
\eqaend
We now add a conformal  anomaly term. Various possibilities exist
 depending on the choice of the measure, and again to cover a general
 situation we take it to be of the form\footnote{ The form assumed
 here is natural from the point of view of world-sheet general covariance.
  However, if one wishes , it can be reproduced by the replacement
 $ K \rightarrow K + \kappa \Omega^2_{,\phi}$ and
 $ q \rightarrow q + \kappa \Omega $. Furthermore, replacements of this
 type can produce a variety of forms for the anomaly term. }
\eqabegin
S_{anom} &=&{\kappa \over \gamma^2} \int d^2x \sqrt{-\ghat}\,\,
  \Biggl(\ghat^{\mu\nu} \delmu(\rho+\Omega(\phi))\delnu
 (\rho+\Omega(\phi))\nn\\
 & & + (\rho+\Omega(\phi))\hat{R}\Biggr) \comma
\eqaend
 allowing an arbitrary function $\Omega(\phi)$.
 The dependence of  $\kappa$ on the number $N$ of matter fields,
 to be explicitly incorporated later, may differ depending again
 on the choice of the measure \cite{St,RST}, but  here it need not
 be specified. With this term added,
 the resultant action can be identified with a sigma model discussed
 in the previous section with
\eqabegin
(X^1, X^2) &=& (\phi, \rho) \comma \\
Q &=& q + \kappa (\rho + \Omega) \label{Q} \comma \\
V &=& 2\rho+v+2\omega \label{V} \comma \\
G_{ij} &=& \matrixii{K(\phi)+2q_{,\phi}\omega_{,\phi}+
\kappa (\Omega_{,\phi})^2}{Q_{,
\phi}}{Q_{,\phi}}{\kappa} \period
\eqaend
It is useful to note that the contravariant vector $Q^{,k}$ has a
 particularly simple form, namely $Q^{,k}=(0,1)$,  due to the fact
 that $Q_{,k}$ is precisely the same as the second column vector
 of $G_{ij}$. Using this, one can easily check that the conformal
 invariance conditions  $Q^{,k}V_{,k} =2$ and $\nabla_i
 \nabla_j Q =0$ are satisfied for any choice of the functions above.
 This is understood as follows: Since $ S_{anom} $ has the form
 maintaining the world-sheet covariance,
the total action $ S + S_{anom} $ also has this covariance.
Then the traceless condition becomes identical with the invariance
with respect to the variation of the conformal factor.
Thus  $ (T^{DG})^\mu_\mu = 0 $
 is satisfied automatically by the equation of motion.
\parsmallskip
Also, inner product of any target-space vector with $Q^{,k}$ is
 easy to compute. In particular, we note
\eqabegin
 Q_{,k} Q^{,k} &=&Q_{,\rho} =  \kappa \period \label{eqn:QQ}
\eqaend
%
\subsection{ Condition for Solvability }
In the previous section, we have seen that when $\nabla_i\nabla_j V
 =0$ is satisfied a powerful symmetry exists and it leads to the
 solvability of the model.  We now analyze this condition
 in detail. \parsmallskip
{}From the structure of $G_{ij}$ it is easily to check that the only
non-vanishing components of the Christoffel symbols are
\eqabegin
 \Gamma^{\phi}_{\phi\phi} &=&  {\del_\phi |G| \over 2|G|} \comma \\
 \Gamma^{\rho}_{\phi\phi} &=& {1\over 2|G|}
\left( 2G_{\phi\phi}G_{\phi\rho,\phi} -G_{\phi\rho}G_{\phi\phi,\phi}
 \right)\comma
\eqaend
and that  all the components of $\nabla_i\nabla_j V
 =0$, except $\nabla^2_\phi V=0$, are automatically satisfied.
This non-trivial condition reads
\eqabegin
 \nabla_\phi^2 V &=& \del_\phi\Vphi -\half \Vphi{\del_\phi |G|
 \over |G|}\nn\\
&& - {1\over |G|}\left( 2G_{\phi\phi}G_{\phi\rho,\phi}
-G_{\phi\rho}G_{\phi\phi,\phi} \right) =0\period
\eqaend
After a straightforward but somewhat tedious calculation, this becomes
\eqabegin
 |G|\nabla_\phi^2 V &=& \qphi^3 \del_\phi \left( {K\over \qphi^2}
 -{\vphi \over \qphi}\right) \nn\\
 && + \kappa W_{,\phi\phi}\left( K + 2\qphi u_{,\phi}\right)
 -{ \kappa \over 2} W_{,\phi} \left( K + 2q_{,\phi}u_{,\phi} \right)_{,\phi} \\
&=& 0 \comma
\eqaend
where
\eqabegin
u &\equiv & \omega -\Omega\comma \\
W &\equiv&  v +2u\period
\eqaend
It is convenient to call the expression in the parenthesis of the first
 line a function $a(\phi)$. In other words, we parametrize the
 function $K(\phi)$  by
\eqabegin
 K &=& a \qphi^2 +\qphi \vphi \label{eqn:K} \period
\eqaend
Substitute this back into the above equation and regard all the quantities
 as functions of $q$ instead of $\phi$. Then the equation simplifies to
\eqabegin
 a_{,q}\left( 1-{\kappa \over 2} W_{,q}\right) + \kappa W_{,qq}
\left( a+\half W_{,q}\right) &=& 0 \label{eqn:aW} \period
\eqaend
There are two types of solutions to this equation. \parmedskipn
Type I\ :\quad When the quantity in the first parenthesis does not
 vanish, one can convert this into a differential equation for the function
 $a$ with respect to $y\equiv W_{,q}$.  The general solution can then be
 readily obtained:
\eqabegin
 \kappa a &=&1-\kappa W_{,q} - c( \kappa W_{,q} -2)^2 \\
 &=& (1-4c) (1-\kappa W_{,q}) -c\kappa^2 W_{,q}^2 \comma \label{eqn:sola}
\eqaend
where $c$ is an arbitrary integration constant. The second line suggests
 that $c=1/4$ is a rather special value.  The significance of this value
 is understood when we compute the quantity $\d=V^{,k}V_{,k}$, which
 gives the cosmological constant for the Liouville equation satisfied
 by  $V$.  With the parametrization (\ref{eqn:K}), $\d$
 can be expressed as
\eqabegin
 \d &=& {4a + \kappa W_{,q}^2 \over \kappa ( a+ W_{,q}) -1}
\label{eqn:genVV}
\period
 \eqaend
Substituting (\ref{eqn:sola}) into this, we get
\eqabegin
 \d &=& {4c-1 \over \kappa c} \period \label{eqn:VV}
\eqaend
Thus we see that {\it $c=1/4$ is precisely the value at which the field
 $V$ becomes a free field.}  As we shall see, all the existent models
 fall into this category.  \parmedskipn
Type II\ : \quad In the special case where $\kappa W_{,q}=2$,
 the equation  (\ref{eqn:aW}) is satisfied with an arbitrary $a(\phi)$ since
$W_{,qq} =0$.  The general formula (\ref{eqn:genVV}) gives in this
 case
\eqabegin
  \d &=&{4\over \kappa} \period
\eqaend
Thus models of this category are ill-defined without the anomaly term
 and we shall not consider them further.
\subsection{ Further Analysis of  Type I Solution}
As was demonstrated in Sec.2, the defining equation for $F$, namely
 $F_{,k} = - \sqrt{-|G|}\, \ep_{kl} V^{,l}$ is integrable. For the type I
 solution, integration can indeed be carried out explicitly. It is most
 convenient to cast the result in the form
\eqabegin
 F &=& {1\over 2\sqrt{c}} \left( V  -2c \d r \right) \comma\label{eqn:defF}
\eqaend
where
\eqabegin
 r &\equiv & q-{\kappa \over 2} W = q-{\kappa\over 2}(v+2\omega
 -2\Omega) \period\label{eqn:defr}
\eqaend
This quantity $r(\phi)$ will be useful in analyzing the
space-time structure as it can be readily expressed  in terms
 of the free fields $\psi$ and $\chi$.  Substituting  (\ref{eqn:defchi}) and
 (\ref{eqn:solY}) into (\ref{eqn:defF})  we get
\eqabegin
 r &=& -\chi -{1\over c\d} \ln \left( 1+{\Lambda \d \over 16} AB\right)
 \comma  \label{eqn:rpsichi}
\eqaend
where we have used
\eqabegin
 c_V &=& 2\sqrt{c}  = \left(1-{\kappa \d \over 4}\right)^{-1/2}
\period \label{eqn:cV}
\eqaend
The latter equation is obtained by noting that Eq.(\ref{eqn:defF}) can also be
written in the form $F=c_V(V -
\half \d Q)$ as in (\ref{eqn:FVQ})
and by using Eqs.(\ref{Q}),(\ref{V}) and (\ref{eqn:VV}).
It conforms to the general formula
 (\ref{eqn:cVQ}) derived towards the  end of  Sec.2 . \parsmallskip
Now let us examine the type I solution  (\ref{eqn:sola})  more closely.
Using  $K =\qphi^2(a+v_{,q})$ and $u=\omega -\Omega$,
 it can be rewritten as a relation between various functions  appearing
 in the action:
\eqabegin
 c(\kappa \vphi +2(\kappa u_{,\phi}-\qphi))^2 &=& \qphi^2 -2\kappa
 \qphi u_{,\phi} -\kappa K \period
\eqaend
Thus we can solve for one of the functions if the others are known.
 For instance, if we regard the potential function $v(\phi)$ as unknown,
 we can easily solve for it to get
\eqabegin
  v &=& {2\over \kappa}(q-\kappa u)  \nn\\
 && \quad \pm {1\over \kappa \sqrt{c}}\int^\phi d\phi \sqrt{ \qphi^2 -\kappa K
 -2\kappa \qphi u_{,\phi}} \period \label{eqn:v}
\eqaend
We shall shortly describe an application of this formula.
\subsection{ Existing Models }
We now show how the various solvable models so far proposed
 are naturally  described in our unified scheme. It turns out that they
 can be classified into two categories depending on whether $W_{,qq}$
 in Eq.(\ref{eqn:aW}) vanishes or not.  \parsmallskip
First, the models with non-vanishing $W_{,qq}$ can  all be
 regarded as  special cases of the one analyzed by de Alwis
\cite{Al}, which is specified by
\eqabegin
 \omega(\phi) &=& \Omega(\phi)  =0\comma \\
 K(\phi) &=& 4\e^{-2\phi} ( 1+h(\phi)) \comma \\
 Q_{,\phi} &=&q_{,\phi}=  -2\e^{-2\phi} ( 1+\bar{h}(\phi)) \comma
\eqaend
where $h(\phi)$ and $\bar{h}(\phi)$ describe possible deviations from the
 CGHS model.  Integrating the last equation, we get
\eqabegin
 q &=& \e^{-2\phi} -2\int^\phi \e^{-2\phi}\, \bar{h}\,  d\phi \period
\eqaend
Substituting these functions into (\ref{eqn:v}) with a choice
of the $ + $ sign immediately yields
 the potential function
\eqabegin
v(\phi) &=& {2q \over \kappa} +{2\over \kappa \sqrt{c}}
 \int^\phi  \ e^{-2\phi} \sqrt{(1+\bar{h})^2  -\kappa \e^{2\phi}(1+h)}
d\phi \nn
\eqaend
For the special value $c=1/4$ discussed previously, this is precisely
 the form  obtained in
\cite{Al} by  use of complicated field transformations.  (Our sign
 convention for $\kappa$ is opposite to de Alwis's.)  \parsmallskip
 For certain choices of $h$ and $\bar{h}$, the integrals can be explicitly
 performed. For the model of Bilal and Callan \cite{BC}, $h=\bar{h}=0$ and
 one gets
\eqabegin
q&=& \e^{-2\phi} \comma \\
v &=& {2\over 1+Z} +\ln {1+Z \over 1-Z} + const. \comma\\
Z &\equiv &  \sqrt{1-\kappa q^{-1}} \period
\eqaend
As for the RST model \cite{RST} ,   $h=0, \bar{h}=(\kappa /4)
 \exp(2\phi)$ and one has
\eqabegin
 q &=& \e^{-2\phi} -{\kappa \over 2}\phi \comma\\
 v &=& -2\phi \period
\eqaend
\indent
 On the other hand, for the model treated in \cite{HKS}-\cite{KS2},
 \cite{VV},
 $W_{,qq}$ vanishes and  hence one has an additional relation
\eqabegin
 W &=& v+2(\omega -\Omega) = 2b q + const. \comma\nn
\eqaend
where $b$ is a constant.  The model is then specified by
\eqabegin
 \Omega &=& \kappa =0 \comma \label{Omk}\\
q &=& \e^{-2\phi} \comma \qquad K = 4q \comma \\
v &=& \ln q \comma \\
\omega &=& q-\half \ln q \qquad (b=1) \label{om}\period
\eqaend
(It can be easily checked that $\d=0$ (\ie $c=1/4$)  and the
 general formula (\ref{eqn:v})
 is still valid  with the limiting procedure $\kappa \rightarrow 0$.)
 \parsmallskip
In the discussions above, which evidently demonstrate
   the  generality of our scheme, the value of $c$
 turned out to be  invariably
 equal to  $1/4$ for all the models treated. There is  in fact a simple way
 to understand it.   Suppose we take a restrictive point of view that
  the potential function $\e^V$ in the  action (\ref{Sflat})  be a marginal
 deformation as in the procedure of  David, Distler and Kawai \cite{DDK}.
 Then it should be a  $(1,1)$ operator and  one can easily show that this
requirement leads to the value $c=1/4$.
\section{ Space-time Structure of the Models }
\sectionnumbering
In the preceding sections, we have systematically analyzed a general
 class of solvable dilaton gravity models and  shown that such
 systems can be invariably described by a set of free fields $\psi$
 and $\chi$. We now add free massless matter fields to the system
 and study the space-time structure thereby  induced.
\subsection{Energy-Momentum Tensor Constraints}
We shall consider the situation where left-going
 matter fields are sent in from the past null infinity, which is described
 by an energy-momentum tensor $T^f_{++}$. Further, to make
 the  analysis  simple, we   take the
 $\psi=0$ gauge.  Then, recalling the form of $T_{++}^{DG}$
 given in Eq.(\ref{eqn:EMpc}),  with   $c_V = 2\sqrt{c}$,
the energy-momentum constraint becomes
\eqabegin
 \delplus^2\chi -c\d (\delplus \chi)^2 &=& T^f_{++} \period
\label{eqn:EMconst}
\eqaend
Depending on the behavior of the model, it is sometimes necessary
 to add a contribution of a boundary function \cite{CGHS} if one wishes
 to realize  asymptotic  flatness in the region which is in the causal
 past to the injection of the matter.
To make the analysis transparent, we   will  first  focus on the cases
 where such  a  function  is not necessary and will briefly discuss  the
 remaining cases later. \parsmallskip
Now when
 the cosmological constant $\d$ is non-vanishing, Eq.(\ref{eqn:EMconst})
 is non-linear as well as inhomogeneous and it is difficult to
 solve it  for general $T^f_{++}$.  However, for a shock-wave
 configuration of the form
\eqabegin
 T^f_{++} &=& T_0\delta(\xplus -\xplus_0)\comma \label{shwave} \\
& & T_0 > 0 \comma \qquad  \xplus_0 > 0 \comma
\eqaend
the exact solution can be obtained. \parsmallskip
We shall now solve it with the boundary condition $\chi = 0$ for
$\xplus < \xplus_0$. (More general choice is possible, but it does not
 appear to give qualitatively different results. )
  Let us write $x=\xplus -\xplus_0$ and set
$\del_x \chi(x) = \theta(x) w(x)$.  Then the equation becomes
\eqabegin
 \delta(x) w(0) + \theta(x) w'(x) - c\d  w(x)^2 \theta(x)
 &=& T_0 \delta(x) \period
\eqaend
Thus $w(x)$ must satisfy
\eqabegin
 w(0) &=& T_0\comma \\
 w'-c\d w^2 &=& 0 \period
\eqaend
The solution is
\eqabegin
 w(x) &=& {T_0 \over 1-c\d  T_0 x} \period
\eqaend
Integrating once again, we obtain the desired solution
\eqabegin
 \chi(\xplus -\xplus_0) &=& {-1\over c\d} \ln Z \comma
\eqaend
where
\eqabegin
Z &\equiv &  1-c\d  T_0 \theta(\xplus -\xplus_0) ( \xplus -
 \xplus_0) \period \label{eqn:Z}
\eqaend
%
\subsection{ Formula for the Curvature Scalar}
We are  now  ready to examine the behavior of the scalar
 curvature. The original metric can be written in the form
\eqabegin
 g_{\mu\nu} &=& \e^{2(\rho +\omega)}\eta_{\mu\nu}
 =  \e^{V-v}\eta_{\mu\nu} \period \label{eqn:metric}
\eqaend
Using the Liouville equation of motion for $V$ and its solution
 given in (\ref{eqn:solV}) -- (\ref{eqn:AB}), the curvature scalar
 in the $\psi=0$ gauge can be written as
\eqabegin
 R &=& -\e^{v-V}\Box ( V-v)
 = \e^v\left( -\half \Lambda \d + Y^2 \Box v \right) \label{eqn:R} \comma
\eqaend
where
\eqabegin
 Y &=& 1+{\Lambda\d \over 16} \xplus\xminus \period\label{eqn:Y}
\eqaend
In order for the field $V$ to be real, $Y$ must be positive. Rather than
 enumerating all possibilities, we shall take $\d
 \le 0$ so that the quadrant $0< \xplus < \infty, \,
 -\infty < \xminus < 0$, which is considered for the existing
 models, is included in the allowed region.  \parsmallskip
Let us recall the quantity $r(\phi)$ given in (\ref{eqn:rpsichi}).
 Since $\chi$ and $Y$ are already known, we immediately
 find the explicit form of $r(\xplus, \xminus)$:
\eqabegin
 r &=&{1\over c\d } \ln \left( {Z \over Y}\right) \comma
\eqaend
where $Z$ and $Y$ are as  in (\ref{eqn:Z}) and (\ref{eqn:Y}).
 It is important to
 note that the behavior of $r$ is completely fixed {\it independently}
of the choice of various functions $q(\phi)$, $K(\phi)$, $v(\phi)$,
$\omega(\phi)$ and $\Omega(\phi)$.
Therefore in analyzing
 $R$ it is convenient to regard $v$  as  a function of $r$. \parsmallskip
Before deriving a general formula for $R$, it is instructive to look at the
 behavior of $r(\xplus, \xminus)$ for  several regions, which we shall
use shortly. First,
 as $\xplus \rightarrow 0$, $r$ tends to a $\d$-independent
 expression
\eqabegin
 r &\longrightarrow & {-\Lambda\over 16c} \xplus\xminus \period
 \label{eqn:xpzero}
\eqaend
 On the other hand, for the asymptotic region of large $\xplus$, the
 behavior of $r$ depends crucially on whether $\d$ vanishes or not.
 For models with $\d=0$, we have
\eqabegin
 r &\longrightarrow   & -{1\over 4}\xplus (\Lambda \xminus + 4T_0)  \comma
\eqaend
whereas  for non-vanishing $\d $, $r$ tends to a finite function of
$\xminus$ given by
\eqabegin
 r &\longrightarrow & {1\over c\d} \ln {16cT_0 \over -\Lambda\xminus}
 + \calO(1/\xplus) \period \label{eqn:rnzd}
\eqaend
Another important region is the one where $r$ vanishes, which occurs
 for $Z=Y$. There are two such regions. One occurs after the shock-wave
 has passed, namely for  $\xplus> \xplus_0$, along the hyperbolic line
\eqabegin
 \xplus(\Lambda\xminus + 16cT_0) &=&
16cT_0 \xplus_0 \label{eqn:rvanish} \comma
\eqaend
which is independent of $\d$.  We shall see that for many models,
 a black hole  singularity is located along this line.  The other region
 of vanishing $r$ is along the asymptotic line $\xplus\xminus \sim 0$
before the shock-wave has passed.  \parsmallskip
Let us now go back to the evaluation of $R$.  Regarding $v=v(r)$,
 Eq.(\ref{eqn:R}) becomes
\eqabegin
-R &=& \e^v\left[ \half \Lambda\d +4 Y^2 \left(
v_{,r} \delplus\delminus r +v_{,rr} \delplus r \delminus r  \right)
 \right] \period
\eqaend
Derivatives of $r$ are easily computed and  we obtain a general formula:
\eqabegin
 -R &=& \Lambda\e^v \Biggl\{ \half \d - {1\over 4c}v_{,r}
+ {1\over (8c)^2 Z} \Biggl[ \xplus(\Lambda\xminus + 16cT_0
\theta(\xplus-\xplus_0))  \nn\\
 & & \quad
+c\d \Lambda T_0 \xplus_0 \theta(\xplus-\xplus_0)\xplus\xminus \Biggr]
 v_{,rr} \Biggr\} \label{eqn:genR}
\eqaend
In this form one can readily take  the limit $\d \rightarrow 0$,
 in which the expressions for  $r$ and $R$ simplify considerably  to
\eqabegin
r &=& -T_0 \theta(\xplus -\xplus_0)(\xplus -\xplus_0)
- {\Lambda\over 4}\xplus\xminus  \comma \label{eqn:rzero} \\
 R &=&\Lambda \e^v \left( v_{,r} +(r-T_0\xplus_0\theta(\xplus -\xplus_0))
v_{,rr} \right) \label{eqn:Rzero} \period
\eqaend
\subsection{New Models}
We now make use of the formula for $R$ and explore models with
 reasonable physical behaviour. Because of the generality of our
 scheme, large possibilities exist and we shall have to limit our
investigation to only a certain class of models satisfying a number
 of simplifying features. First, we assume
 $\omega =\Omega =0$ as in the original CGHS model and concentrate
 on the cases of $\kappa \geq 0 $.
 Next it is  natural to suppose that the potential function $v(\phi)$ is
  independent of the number of matter fields  and hence on $\kappa$.
 $q(\phi)$, on the other hand,  may or may not depend on $\kappa$
 but we shall consider the  $\kappa$-independent cases. This means that
 we can regard $v$ as a $\kappa$-independent function of $q$ and
 we have the relation $r=q -(\kappa/2)v(q)$ from (\ref{eqn:defr}).
It would be most
 convenient if we can solve this relation to get  $v(r)$, but it is in general
 difficult.  Thus we treat $v$ as $v(q)$ and substitute in the formula
 for $R$ the relations
\eqabegin
 v_{,r} &=& {v_{,q} \over 1-{\kappa \over 2} v_{,q}} \comma \qquad
 v_{,rr} =  {v_{,qq} \over \left(1-{\kappa \over 2} v_{,q}\right)^3 }
 \label{vq}
\eqaend
Our analysis will be such that the specific form of $q(\phi)$ will not be
 needed.
\subsubsection{Models with $\d=0$ }
Let us begin with  models with  $\d=0$.  First we look at the behavior of
 $R$ for the asymptotic region of large $\xplus$, where $r\rightarrow
+\infty$.  To motivate our search for a reasonable $v(q)$, recall the
CGHS model, where $q=\e^{-2\phi}$ is a measure of the inverse
 coupling strength.  It would then be natural to require that $r\rightarrow
 +\infty$ corresponds, at least for sufficiently large  $|\xminus |$,
 to the weak coupling region with  large positive $q$.  Then
 we wish to have
 $ r \sim q^a$ with positive $a$ in that region.  This prompted  us to
 investigate the following two cases:
\eqabegin
&& (i)\quad  v(q)= -\alpha q^{1+\beta}, \quad (\alpha>0\comma
\beta >-1)\comma
 \qquad  r = q + (\kappa \alpha / 2) q^{1+\beta} \\
&& (ii)\quad  v(q) =-\alpha \ln q \comma
 \qquad  r= q+  (\kappa \alpha / 2) \ln q
\eqaend
Although we are focusing only on a particular asymptotic region,
 these forms for $v(q)$ are rather generic  and useful for the investigation
 of  other  regions as well. \parsmallskip
%
Let us begin with the type (i) potential. As we wish to  allow for non-integral
 value of $\beta$ and  in fact  for  just such a case we shall find
 interesting space-time structure,   the range of $q$ is restricted to
 $q\ge 0$ in order for $ v(q) $ to be real.  This  implies  that $r$
 should also be non-negative. Recalling the expression of $r$  given in
 (\ref{eqn:rzero})  we shall therefore concentrate on (a part of) the quadrant
$0< \xplus < \infty, \,  -\infty < \xminus < 0$.
The exact form of $ R $ is given by
\eqabegin
 R_{(i)} & =& -\alpha \Lambda \e^{-\alpha q^{1+\beta}}
 \left[
{    (1+\beta) q^\beta   \over   1+{\kappa \alpha\over 2}(1+\beta) q^\beta   }
\right. \nn\\
&& \left. +{  \beta (1+\beta)   \over
 \left(   1+{\kappa \alpha\over 2}(1+\beta) q^\beta   \right)^3     }
 \left(
q^\beta \left(    1+ {\kappa \alpha\over 2}q^\beta    \right)
 -T_0\xplus_0 q^{\beta-1} \theta(\xplus -\xplus_0)  \right)
   \right] \period  \label{eqn:Rpower}
\eqaend
{}From this expression, we can read off the behavior of $ R $.
First, $ R_{(i)} $ vanishes in the limit $ \xplus \to \infty $
for positive $ \alpha $ because
large $ r $ corresponds to large $ q $.
Next we study the behavior as $\xplus \rightarrow 0$ and also along
 the hyperbolic line (\ref{eqn:rvanish}).  As was pointed out previously,
$r$ tends to vanish in both of these regions.
For  $\xplus \rightarrow 0 $,
$q$ also vanishes and this leads to the vanishing
 of $R$ for any positive  $\beta$ since  the term proportional to the
 $\theta$-function describing the shock wave is absent in  this region.
 In contrast,  for $\xplus  > \xplus_0$,  a curvature singularity may develop
 along $r=0$ line.
 Indeed for a fractional value $1 > \beta > 0$, we  get a singularity
 of strength $R \sim q^{\beta -1}$  precisely due to the shock wave
 and by drawing a Penrose diagram one can easily identify this as
 a black hole singularity accompanied by an event horizon along
 $\xminus = -16cT_0/\Lambda$. \parsmallskip
As for the type (ii), the exact form of $ R $ becomes
\eqabegin
 R_{(ii)} &=& \Lambda
{  \alpha q^{-\alpha} \over q+{\kappa \alpha  \over 2}   }
 \left[
  -1+ {q   \over   \left( q+{\kappa \alpha \over 2}\right)^2}
 \left( q+ {\kappa \alpha \over 2} \ln q -T_0 \xplus_0
\theta(\xplus -\xplus_0)\right)
\right] \period \label{eqn:Rlog}
\eqaend
In this case, the behavior of $ R $ is very different depending on
whether $ \kappa $ vanishes or not.  For  $ \kappa = 0 $,
$ R_{(ii)} $ becomes proportional to the $ \theta $-function.
Then, noting that $ r = q $ in this case,
we see that $ R_{(ii)} $ vanishes identically for $ \xplus < \xplus_0 $.
For
$ \xplus > \xplus_0 $, the scalar curvature
tends to vanish as $ \xplus \to \infty $ for $ \alpha > -2 $, and
a curvature singularity due to  the shock wave forms along the
$ r = 0 $ line.
Note that $ v(q) $ is well-defined only in the quadrant
 $0< \xplus < \infty, \,  -\infty < \xminus < 0$.
\parsmallskip
Next, let us consider the case with non-zero $ \kappa $.
 For positive $\alpha$, $r $  goes like  $q$ for large $r$ and
  by  simple power
 counting we see that $R_{(ii)} \rightarrow 0$. When $\alpha <0$,
 large $r$ corresponds both to large and small $q$.  For small $q$,
 again by power counting we ascertain the vanishing of $R_{(ii)}$.
 For large $q$, the asymptotic behavior
is $R_{(ii)} \sim q^{-(2+\alpha)}$
 and it vanishes for $\alpha > -2$. Thus we find that these models do
 have asymptotically flat regions for large  $\xplus$
 provided $\alpha > -2$. \parsmallskip
 As for the behavior as $ \xplus \to 0 $ and along the hyperbolic line
$ r = 0 $, the situation again depends very much
 on the value of $\alpha$.  First for $\alpha< 0$, $ r(q) $ has a minimum
$ r_{min} $ at $ q = -\kappa \alpha /2 $,
and just at this point the denominators of
$ v_{,r} $ and $ v_{,rr} $ vanish. Therefore a curvature singularity
develops along the line $ r = r_{min} $ before $ \xplus \to 0 $ or $ r = 0 $
is reached. \parsmallskip
   On the other hand, for positive $\alpha$,  $r=0$
 corresponds to a non-vanishing value  $q=q_c$,
 defined as the solution of
 the equation $q_c + (\kappa \alpha/2)\ln q_c =0$. It is easy to see that
 $0<q_c <1$ and it vanishes as $\kappa \rightarrow 0$.  Then
for both of the above regions $R$ becomes constant and takes the form
\eqabegin
 R &\longrightarrow & -{\alpha \Lambda \over q_c^\alpha
 \left( q_c + {\kappa \alpha \over 2} \right) } \: \delta_{\kappa, 0}
  \qquad ( \xplus \rightarrow 0 ) \comma
 \\
 R&\longrightarrow & -{\alpha \Lambda \over q_c^\alpha
 \left( q_c + {\kappa \alpha \over 2} \right) }
 \left( (1- \delta_{\kappa,0}) +
 {T_0 \xplus_0 q_c \over \left(q_c + {\kappa \alpha \over 2}\right)^2 }
\right)
 \qquad ( \xplus >\xplus_0, r\rightarrow 0 ) \period
\eqaend
This means that as $\xplus \rightarrow 0$ the space-time is anti de Sitter
 instead of flat and at the same time the curvature singularity along
 $r=0$ disappears.
 Although this appears  interesting, there is actually a problem:
For $ \kappa \neq 0 $, $ v(q) $ becomes well-defined
even in the region for  negative $ r $, and as  $ r \to - \infty$ the curvature
$R$ is seen to diverge.
\parsmallskip
 Combining the results obtained above, we can construct an interesting
 new model. It is obtained by smoothly
 joining the  type (i)  potential for the region $\xplus \lsim \xplus_0$
 and the type (ii) potential for the remaining region
$\xplus  \gsim \xplus_0$.
An example is given by
\eqabegin
 v(q) &=& {-\alpha q^{1+\beta} +  q^2\ln q \over 1+q^2}
 \comma \qquad (\alpha>0, \ 0<\beta <1 )
\eqaend
 The allowed region, where  $ v(q) $ is well-defined, is
again seen to be
 the quadrant $0< \xplus < \infty, \,  -\infty < \xminus < 0$.
\parsmallskip
The model so obtained has the following attractive properties :
(a) a  black hole singularity forms along the
 space-like hyperbolic line (\ref{eqn:rvanish}),
(b) the space-time is flat for both $\xplus \rightarrow
 \infty$ (outside the horizon) and $\xplus \rightarrow 0$, and
(c)  Hawking radiation exists.
 Let us see how they come about. First note that
 for large $q$ the potential is dominated by the log term, while for
 small $q$  the power term prevails and in both regions $r \sim q$ holds.
 Thus from the previous analysis the properties (a) and (b) are guaranteed.
 To see that Hawking radiation exists in
 this model, recall the form of the metric.
In the $\psi=0$ gauge with $\d=0$, the form given in  (\ref{eqn:metric})
 reduces to  $g_{\mu\nu} = \e^{-v}\eta_{\mu\nu}$.  For $\xplus \rightarrow
 0$,  $v(q)$ vanishes and hence $x^\pm$ are  the manifestly
 flat coordinates.  On the other hand, for $\xplus \rightarrow \infty$,
 we have $q \sim r \sim -{1\over 4}\xplus (\Lambda \xminus + 4T_0)$
 and the line element becomes
\eqabegin
 ds^2 &=& {4d\xplus d\xminus \over \xplus (\Lambda \xminus + 4T_0)}
\period
\eqaend
Thus to get  manifestly flat coordinates  we must make a familiar
 exponential type conformal transformation,  and  by standard arguments
 this leads to  Hawking radiation.
%
\subsubsection{ Models with a boundary function }
Let us now make a brief discussion on the possibility of adding a boundary
 function $\kappa t_{++}(\xplus)$ to $T^f_{++}$ in (\ref{eqn:EMconst})
 in order to realize a flat region for $\xplus < \xplus_0$ when the anomaly
 term is included.
 Originally \cite{CGHS}, possible existence of a boundary function
 was  inferred from the covariant conservation law $\nabla_\mu T^\mu_\nu
 =0$ together with the anomaly equation $g^{\mu\nu}T_{\mu\nu}
 \propto \kappa R$ \cite{CF}.  For us, a slightly different view point
 will be more convenient.  Quantum mechanically,
 $T_{\mu\nu}$
 in these equations must be understood as the expectation value,
 $\langle \hat{T}_{\mu\nu} \rangle_g$, of the operator $ \hat{T}_{\mu\nu}$
 in a curved space-time described by a metric $g_{\mu\nu}$.  More
 precisely, as the form of the anomaly equation dictates,
$ \hat{T}_{\mu\nu}$ must be  defined such that for a flat region  described
 by  manifestly flat metric $g^{\mu\nu}=\eta^{\mu\nu}$  (in  a  coordinate
 system we call $\sigma^\mu$)
   $\langle \hat{T}_{\mu\nu}(\sigma) \rangle_\eta$  vanishes.  If one
 wishes to use  a  different coordinate system, call it $x^\mu$,
related to $\sigma^\mu$ by  a conformal transformation, one must make a
 re-normal-ordering and this can give rise to a  boundary function.
 Since  $R$ still vanishes in $x^\mu$,  one can turn the argument
 around and use $R=0$ (in an appropriate region) as the equation for
  determining the boundary  function.  In what follows, we shall carry
 out this procedure explicitly.
\parsmallskip
For simplicity, we shall limit ourselves  to $\d=0$ case with $\omega
 =\Omega=0$ only.
In this case,  the curvature,  the energy-momentum constraint
for $\xplus < \xplus_0$ and
 the expression for  $\chi$  take the following forms:
 \eqabegin
 R &=& \e^v \Box v \label{Rv} \comma \\
 \delplus^2 \chi &=&\kappa t_{++} \comma \\
\chi &=& -{\Lambda \over 4}\xplus\xminus - r
 = -{\Lambda \over 4}\xplus\xminus -q + {\kappa \over 2}v \comma
 \label{chir}
\eqaend
where the last equation is obtained from  (\ref{eqn:defr}) and
(\ref{eqn:rpsichi}).  Suppose we want  $R$ to
 vanish identically in the above region. Then  $\Box v =0$ must hold
 and we may write $v = v^+(\xplus) + v^-(\xminus)$.
 Putting this into  the expression for $\chi$ and applying $\delplus^2$,
 we get
\eqabegin
 \delplus^2 \chi &=& -\delplus ^2 q + {\kappa \over 2} \delplus^2 v^+
 \period
\eqaend
On the other hand, from $\Box \chi =0$  it follows that  $\delplus\delminus
 q = -{\Lambda \over 4}$ and we get
\eqabegin
 q &=& -{\Lambda \over 4}\xplus\xminus +\kappa  p(x) \comma
 \label{eqn:qform} \\
 p(x) &=& ( p^+(\xplus) +p^-(\xminus) )\comma
\eqaend
where $p^\pm$ are arbitrary functions of the specified variables.
{}From these equations, we get the expression for the boundary function
 $t_{++}$ as
\eqabegin
 \kappa t_{++} &=&\delplus^2\chi = -\kappa \delplus^2 p^+
 + {\kappa  \over 2}\delplus^2 v^+ \period
\eqaend
With the shock wave, $\chi$ is then solved as
\eqabegin
 \chi &=& \kappa \left( -p(x) + \half v(x) \right)
  + T_0 (\xplus -\xplus_0)\theta  (\xplus -\xplus_0)\comma
\label{eqn:solchi}
\eqaend
where $ p^-(\xminus) $ and $ v^-(\xminus) $ are arbitrary functions.
\parsmallskip
When $v$ is specified as a function of $q$, as we have been assuming,
 the form of $v(q)$ satisfying $\Box v=0$ is actually severely restricted
 because of the special form of $q$ given in  (\ref{eqn:qform}). In such
 a case, the condition $\Box v=0$ can be written as
\eqabegin
 0 &=& \delplus\delminus v = \delplus\delminus q \, v_{,q}
 + \delplus q \delminus q\, v_{,qq} \nn\\
 &=& -{\Lambda \over 4} v_{,q} + v_{,qq}
 \left( {\Lambda \over 4} \xminus -\delplus p^+\right)
\left( {\Lambda \over 4} \xplus -\delminus p^-\right) \period
\eqaend
{}From this the following expression must be a function only of $q$:
\eqabegin
 {\Lambda \over 4} {v_{,q} \over v_{,qq}}
 &=& \left( {\Lambda \over 4} \xminus -\kappa\delplus p^+\right)
\left( {\Lambda \over 4} \xplus -\kappa\delminus p^-\right) \period
\eqaend
It is not difficult to prove that it can happen only for the following two
 cases:
\eqabegin
 (i)\qquad  q+c&=& -{\Lambda \over 4} (\xplus + a)( \xminus+b)
\comma \nn\\
 v &=& - \alpha \ln (q +c)+ \beta \comma \\
\kappa p^+ &=& -{\Lambda \over 4} b\xplus + const. \comma \nn\\
v^+ &=& - \alpha \ln (\xplus + a) + const.\comma \nn\\
 (ii)\qquad q+c &=&{\Lambda \over 4}
(a\xplus - {1\over 2a}\xminus+b)^2\comma \nn\\
v &=& \alpha \sqrt{q+c}+\beta \comma \\
\kappa p^+ &=& {\Lambda \over 4} \left( a^2 (\xplus)^2 + 2ab \xplus
 \right) + const. \comma \nn\\
 v^+ &=& \alpha \sqrt{\Lambda /4} a\xplus + const. \comma \nn
\eqaend
where $a, b, c, \alpha$ and $\beta$ are constants.
\parsmallskip
The form of $r$ when a left-going matter shock wave is sent in
 from the past null infinity can then be readily obtained
 from (\ref{eqn:solchi}) and (\ref{chir}) for the above two cases:
\eqabegin
 r_{(i)} &=& - \frac{\Lambda}{4} \xplus \xminus + \frac{\kappa}{2} \alpha
             \ln \left( - \frac{\Lambda}{4} \xplus \xminus \right)
             - T_0 ( \xplus - \xplus_0 ) \theta( \xplus - \xplus_0)
             \comma \\
r_{(ii)} &=& \sqrt{\frac{\Lambda}{4}} ( a \xplus - \frac{1}{2a} \xminus + b )
      \left\{ \sqrt{\frac{\Lambda}{4}} ( a \xplus - \frac{1}{2a} \xminus + b )
               - \frac{\kappa\alpha}{2} \right\} \nn \\
         &&  \qquad \qquad
              -  T_0 (\xplus - \xplus_0) \theta (\xplus - \xplus_0)
              - c - \frac{\kappa}{2} \beta
         \period
\eqaend
Here we have set $ a, b, c, \beta = 0 $ in $ R_{(i)} $ for simplicity.
\parsmallskip
Now from  Eqs.(\ref{Rv}) and (\ref{vq})  the exact form
of $ R $ for these cases is obtained in the same way as in the
previous subsection:
\eqabegin
 R_{(i)} &=& - \alpha \Lambda \frac{ q^{-\alpha}}{q+\kappa \alpha/2}
         \left[ 1 - \frac{q}{( q+\kappa \alpha/2)^2 } \times
         \right. \nn \\
         && \left. \qquad
          \left\{ (-\frac{\Lambda}{4}) \xplus \xminus
            + \kappa \alpha
            - \frac{(\kappa \alpha)^2}{\Lambda \xplus \xminus}
            - T_0 \theta (\xplus - \xplus_0)
            \left( \xplus - \frac{2 \kappa \alpha}{\Lambda \xminus} \right)
            \right\} \right] \comma \\
R_{(ii)} &=& \frac{\alpha}{2}
 \frac{ \Lambda \ e^{\alpha \sqrt{q+ c}+\beta}}{\sqrt{q + c} - \kappa\alpha/4}
          \left[ 1 - \frac{1}{(\sqrt{q + c} - \kappa\alpha/4)^2}
         \left( \sqrt{\frac{\Lambda}{4}}
            ( a \xplus - \frac{\xminus}{2a} + b) - \frac{\kappa\alpha}{4}
            \right) \right. \times \nn \\
          && \left. \qquad \quad
             \left\{ \left( \sqrt{\frac{\Lambda}{4}}
            ( a \xplus - \frac{\xminus}{2a} + b) - \frac{\kappa\alpha}{4}
            \right) - \frac{T_0}{a \sqrt{\Lambda}}
            \theta (\xplus - \xplus_0) \right\} \right] \period
\eqaend
We can easily check that indeed $ R_{(i),(ii)} = 0 $ holds for
$ \xplus < \xplus_0 $ as planned.
\parsmallskip
By making use of these expressions in
the region where $ v(q) $ is well-defined,
It is straightforward to find  the behavior of $ R $ for $ \xplus > \xplus_0 $.
Although there are interesting features in each case,
we omit the details for brevity and comment only on the type (i) case
with $ \alpha < 0 $. \parsmallskip
First for $ \kappa = 0 $, a curvature singularity
develops along the line $ r = 0 $ and $ R $ tends to vanish as
$ \xplus \to \infty $ for $ \alpha > -2 $.
On the other hand, when   $ \kappa(> 0) $ is turned on,
a curvature singularity now forms along
$ r = - (\kappa\alpha/2)( 1 -\ln( - \kappa\alpha/2)) $, which is the
 minimum of
$ r(q) $. In the limit $ \xplus \to \infty $, $ R $ vanishes
for $ 0 > \alpha > -1  $ and diverges for $ \alpha \leq - 1$.
Especially, at the critical value $ \alpha = - 1 $, $ R $ diverges like  $ \ln
\xplus $ and the space-time is essentially the same as that in
 \cite{RST}. Although we have only dealt with a few models,
this   illutrates  how one can incorporate  possible boundary functions and
  widen the possibilities for new models.
\subsubsection{ Models with $\d\ne0$ }
When we allow $\d \ne 0$, it turns out to be quite difficult to find
 a model with asymptotically flat regions.  (This is partly due to the
fact that $r$ stays finite even for $\xplus \rightarrow \infty$, as
 already noted in (\ref{eqn:rnzd}).)  Nevertheless, if we relax this
condition and look for models which are at least asymptotically finite
and at the same time
exhibit a shock-wave-induced black hole singularity, we find that,
 for example, the  power type potential $v(q) = -\alpha q^{1+\beta}$
 examined for $\d=0$ is still viable for the same range
 of $\alpha$ and $\beta$, namely for $\alpha >0, \ 0<\beta<1$, provided
 that $\d<0$. Recall that $ q^{1+\beta} $ is well-defined only in the
quadrant  $0< \xplus < \infty, \,  -\infty < \xminus < 0$.
In this case, as $\xplus$ goes to zero,  $R$ tends to a
positive  constant:
\eqabegin
 R & \longrightarrow & -\half \Lambda \d \period
\eqaend
On the other hand, for  $\xplus \rightarrow \infty$, it goes to
 a finite function of $\xminus$.
Especially in the limit $ -\xminus \to \infty $,
\eqabegin
 R &\longrightarrow & {\alpha \beta(\beta + 1)\Lambda
\over  4c^2 \d q^{1-\beta}
 \left( 1+(\kappa \alpha/2)(1+\beta)q^\beta\right)^3 }
 \left( \frac{-\Lambda\xminus}{16cT_0} \right)^{1-2/(1-4c)}
\comma \\
q^{1+\beta} &=&  \frac{2}{\alpha(1-4c)}
                \ln \left( \frac{-\Lambda \xminus}{16cT_0} \right) \comma
\eqaend
where we have used (\ref{eqn:VV}).
Since $ \d < 0 $ means $ 0 < c < 1/4 $,
this expression tends to vanish as $\xminus
 \rightarrow -\infty$ and we have a flat space at the spatial
 infinity.  As for the black hole singularity, it can easily be seen
 to occur along the $r=0\ ( q=0)$ hyperbolic line in just
 the same way as for the $\d=0$ case previously discussed.
 Admittedly the space-time described by this model is somewhat
 unusual. It is an open question whether one can find models with
 more conventional flat asymptotic behavior by considering
 different $v(q)$  and/or by incorporating  suitable
 boundary functions.
\section{ Discussions }
\sectionnumbering
By observing  that  solvability of  various models of dilaton gravity
 in two-dimensions can be understood  as due to a powerful symmetry,
 we have formulated in a systematic and unified  way how one can
 analyze a rather  general class of such models from the point of
 view of non-linear  sigma model.
This formulation then allowed us to explore a number of new models
 and their space-time behaviors under the influence of the matter shock
 wave.  Although these represent only a small
 subset of models which can be treated by our method, they serve to
 illustrate the general  strategy and the power of our formulation.
\parsmallskip
In this article, we have been using the word \lq\lq solvable" in a broad
 sense;  it means that the model, including the effect of the measure
 as anomaly term,  can be reduced to a collection of  free fields and
 in general a field satisfying the Liouville equation.  As was stated in
 the introduction, to be able to attack the challenging questions
 listed there,  it would be desirable to have a model which is quantum
 mechanically fully solvable, for any number of matter fields. \parsmallskip
 Let us
 briefly discuss what that would require in our formulation. First, in
 this respect, models with a genuine Liouville field are not likely to
 be useful since quantum Liouville dynamics is notoriously difficult,
especially in the operator formalism.  (For its problems,
 see, for example, \cite{KN}). \parsmallskip
 Now restricting to  $\d=0$ case, the
 immediate question is how we can perform the full quantization.
 In a special model with $\kappa =0$ treated in \cite{HKS}, it was shown
  that with an appropriate measure
 the transformation from the original fields to the free fields
 $\psi$ and $\chi$ is a quantum canonical transformation
 and hence it was  justified to use the usual free field quantization.
 It is not difficult to show that this argument can be extended to
 a generalization of that model with non-vanishing $\kappa$.
It is  constructed  in such a way that  the potential term becomes
perturbatively a conformally $ (1,1) $ operator. Specifically the model is
 specified by
 (compare with Eqs.(\ref{Omk})$
\sim $(\ref{om}) )
\eqabegin
 && q = \ e^{-2\phi}\comma  \quad v = \ln q \comma \quad
  \omega = \half \left( 2b q - \ln q \right) \comma \\
&& \Omega = 0 \comma \quad a (\phi) = - \kappa b^2 \comma
    \quad c = \frac{1}{4} \period
\eqaend
As in the model of \cite{HKS},  it is possible to
construct all the physical states explicitly by DDF operators
\cite{DDF} and to further perform the analysis  similar to the one
 presented in  \cite{KS1,KS2}. \parsmallskip
 For a more general class of models treated in this article, similar
 statement has not been demonstrated.
As we shall discuss shortly,
 validity of the  canonicity of transformation has an important bearing
 on the question of the target space general covariance in the string
 theory interpretation of the models. \parsmallskip
Assuming for the moment that free-field quantization is justified, let
 us proceed to examine what other problems we must overcome.
 In that case,  since the energy-momentum tensor has the simple form
\eqabegin
  T^{DG}_{\pm\pm} &=& \del_\pm \psi \del_\pm \chi
 -  \del^2_\pm \chi
 - {\kappa\over 4}\left( (\del_\pm \psi)^2 -2\del^2_\pm \psi \right)
\comma
\eqaend
 the system can be shown to possess quantum conformal invariance.
Thus,  using  the techniques of conformal field theory, it should
 be possible to find all the physical states and so on.
Difficulty appears  when one starts expressing various
 quantities in terms of the free fields.  The quantity directly expressible
 in terms of $\psi$ and $\chi$ is $r$, which for $\d=0$ takes the form
\eqabegin
 r &=& -\chi -{\Lambda \over 4}AB\period
\eqaend
On the other hand, quantities of interest may not be explicitly given
  as functions of $r$.  For example, the conformal
 factor for the metric $\e^{\psi -v}$ involves the potential function $v$.
 Since it is related to $r$ by $r = q -(\kappa /2) v$, if we specify $v$
 as a function of $q$, as we did in the previous section, it is in general not
 possible to get $v(r)$ in a closed form.  A way to circumvent
 this difficulty is, obviously, to specify the model by giving $v(r)$
 directly but then $v(\phi)$ would in general be $\kappa$-dependent.
 Although we did not adopt this point of view in this article,
 it may be worth pursuing  in the fully quantum treatment.\parsmallskip
Finally let us discuss  possible implications of our work when the
 non-linear sigma model is interpreted as a model in string theory.
Since the special symmetry we have imposed can be characterized
 completely in a target space covariant manner, naively all the models
 with equal $\d=V_{,k}V^{,k}$ and $Q_{,k}Q^{,k}$, which are
  target space invariant,  are expected to
 describe the same string theory in  flat background and hence they
 should have identical spectrum. Indeed the energy-momentum tensor
 expressed in terms of the  free fields $\psi$ and $\chi$ depends only
 on these two parameters.  But of course the validity of this
 argument crucially hinges on the nature of the functional measure.
 Only if we can find a general coordinate invariant measure for $X^k$
such that it reduces to a simple measure for the free fields
 $\psi$ and $\chi$ through a canonical transformation, can we
 realize the general coordinate invariance in  string theory.
This is quite similar to the line of thought pursued some years ago
 in \cite{Ve} in a more general setting.
 Since the understanding of the emergence of target space general
covariance in string  theory is an unsolved problem , it would be
 interesting if one can demonstrate it  for our simple models.
\parbigskipn\parbigskipn
{\Large\bf Acknowledgment }
\parbigskip
Y.K. would like to thank  the members,  in particular E. Rabinovici,  of the
 Center of Microphysics and Cosmology
 at the Racah Institute of Theoretical Physics, Hebrew University, for
  cordial hospitality and a generous financial support, where a part of
 this work was carried out.  The research of Y.K. is supported
 in part by the Grant-in-Aid for Scientific Research (No. 06640378) and
 Grant-in-Aid for Scientific Research for Priority Area (No. 06221211),
 while that of Y.S. is partially supported by the JSPS Research
 Fellowship for Young Scientists (No. 06-4391), both from the Ministry
 of Education, Science and Culture.

\end{document}